\documentclass[manuscript]{acmart}
\AtBeginDocument{%
  \providecommand\BibTeX{{%
    \normalfont B\kern-0.5em{\scshape i\kern-0.25em b}\kern-0.8em\TeX}}}

\setcopyright{none}
\renewcommand\footnotetextcopyrightpermission[1]{} % removes footnote with conference information in first column
\setcopyright{none}
\copyrightyear{2023}
\acmYear{2023}
\acmDOI{XXXXXXX.XXXXXXX}

\acmConference[AI Literacy Workshop, CHI'23]{CHI Conference Hamburg 2023}{April 23--28,
  2023}{Hamburg, Germany}

\usepackage{caption}
\usepackage{subcaption}

\usepackage{tikz}
\usepackage{xspace}

\usepackage{xcolor}
\usepackage{ifthen}
\newboolean{include-notes}
\setboolean{include-notes}{true}
\newcommand{\nb}[3]{\ifthenelse{\boolean{include-notes}}{{\colorbox{#2}{\bfseries\sffamily\scriptsize\textcolor{white}{#1}}}{\ \textcolor{#2}{\sf\small\textit{#3}}}}{}}

\newcommand{\amsalgorithm}{\textit{AMS algorithm}\xspace}

\begin{document}

%%
%% The "title" command has an optional parameter,
%% allowing the author to define a "short title" to be used in page headers.
\title[Interdisciplinary Frameworks to Understand Algorithmic Decision-Making]{Applying Interdisciplinary Frameworks to Understand Algorithmic Decision-Making}

%%
%% The "author" command and its associated commands are used to define
%% the authors and their affiliations.
%% Of note is the shared affiliation of the first two authors, and the
%% "authornote" and "authornotemark" commands
%% used to denote shared contribution to the research.
\author{Timothée Schmude}
\email{timothee.schmude@univie.ac.at}
\affiliation{%
  \institution{University of Vienna, Faculty of Computer Science, Research Network Data Science, UniVie Doctoral School Computer Science DoCS}
  \streetaddress{Währinger Straße 29}
  \city{Vienna}
  \state{Vienna}
  \country{Austria}
  \postcode{1090}
}

\author{Laura Koesten}
\email{laura.koesten@univie.ac.at}
\affiliation{%
  \institution{University of Vienna, Faculty of Computer Science, Research Group Visualization and Data Analysis}
  \streetaddress{Sensengasse 6}
  \city{Vienna}
  \state{Vienna}
  \country{Austria}
  \postcode{1090}
}

\author{Torsten Möller}
\email{torsten.moeller@univie.ac.at}
\affiliation{%
  \institution{University of Vienna, Faculty of Computer Science, Research Network Data Science, Research Group Visualization and Data Analysis}
  \streetaddress{Sensengasse 6}
  \city{Vienna}
  \state{Vienna}
  \country{Austria}
  \postcode{1090}
}

\author{Sebastian Tschiatschek}
\email{sebastian.tschiatschek@univie.ac.at}
\affiliation{%
  \institution{University of Vienna, Faculty of Computer Science, Research Network Data Science, Research Group Data Mining and Machine Learning}
  \streetaddress{Währinger Straße 29}
  \city{Vienna}
  \state{Vienna}
  \country{Austria}
  \postcode{1090}
}

%%
%% By default, the full list of authors will be used in the page
%% headers. Often, this list is too long, and will overlap
%% other information printed in the page headers. This command allows
%% the author to define a more concise list
%% of authors' names for this purpose.
\renewcommand{\shortauthors}{Schmude et al.}

%%
%% The abstract is a short summary of the work to be presented in the
%% article.
\begin{abstract}
  We argue that explanations for "algorithmic decision-making" (ADM) systems can profit by adopting practices that are already used in the learning sciences. We shortly introduce the importance of explaining ADM systems, give a brief overview of approaches drawing from other disciplines to improve explanations, and present the results of our qualitative task-based study incorporating the "six facets of understanding" [24] framework. We close with questions guiding the discussion of how future studies can leverage an interdisciplinary approach. 
\end{abstract}

%%
%% The code below is generated by the tool at http://dl.acm.org/ccs.cfm.
%% Please copy and paste the code instead of the example below.
%%
\begin{CCSXML}
<ccs2012>
   <concept>
       <concept_id>10003120.10003121.10003126</concept_id>
       <concept_desc>Human-centered computing~HCI theory, concepts and models</concept_desc>
       <concept_significance>500</concept_significance>
       </concept>
   <concept>
       <concept_id>10003120.10003121.10003122.10003334</concept_id>
       <concept_desc>Human-centered computing~User studies</concept_desc>
       <concept_significance>500</concept_significance>
       </concept>
 </ccs2012>
\end{CCSXML}

\ccsdesc[500]{Human-centered computing~HCI theory, concepts and models}
\ccsdesc[500]{Human-centered computing~User studies}

%%
%% Keywords. The author(s) should pick words that accurately describe
%% the work being presented. Separate the keywords with commas.
\keywords{education/learning, empirical study that tells us about people, interview, qualitative methods}

%% A "teaser" image appears between the author and affiliation
%% information and the body of the document, and typically spans the
%% page.
\begin{comment}
\begin{teaserfigure}
  \includegraphics[width=\textwidth]{sampleteaser}
  \caption{Seattle Mariners at Spring Training, 2010.}
  \Description{Enjoying the baseball game from the third-base
  seats. Ichiro Suzuki preparing to bat.}
  \label{fig:teaser}
\end{teaserfigure}
\end{comment}

%%
%% This command processes the author and affiliation and title
%% information and builds the first part of the formatted document.
\maketitle

\section{Introduction}

"Algorithmic decision-making" systems analyse large amounts of data to support and drive decision-making in public and private institutions~\cite{european_parliament_understanding_ADM_2019}, often affecting high numbers of people. Well-known examples of such "high-risk"~\cite{european_commission_laying_2021} systems can be found in recidivism prediction~\cite{chouldechova2017}, refugee resettlement~\cite{Bansak2018}, and public employment~\cite{scott_algorithmic_2022}. Many authors have outlined that faulty or biased predictions by ADM systems can have far-reaching consequences, including discrimination~\cite{chouldechova2017}, inaccurate predictions~\cite{brown_toward_2019}, and overreliance on automated decisions~\cite{allhutter_bericht_ams-algorithmus_2020}. Therefore, high-level guidelines are meant to prevent these issues by pointing out ways to develop trustworthy and ethical AI~\cite{thiebes_trustworthy_2021, floridi_ai4peopleethical_2018}. However, practically applying these guidelines remains challenging, since the meaning and priority of ethical values shift depending on who is asked~\cite{jakesch_how_2022}. 

Recent work in Explainable Artificial Intelligence (XAI) thus suggests equipping individuals who are involved with an ADM system and carry responsibility---so-called "stakeholders"---with the means of assessing the system themselves, i.e. enabling users, deployers, and affected individuals to independently check the system's ethical values~\cite{langer_what_2021}. Arguably, a pronounced \textit{understanding} of the system is necessary for making such an assessment. While numerous XAI studies have examined how explaining an ADM system can increase stakeholders' understanding~\cite{shulner-tal_enhancing_2022, szymanski_visual_2021}, we highlight two aspects that remain an open challenge: i) the amounts of resources needed to produce and test domain-specific explanations and ii) the difficulty of creating and evaluating understanding for a large variety of people. % and iii) the need to adapt the explanation to changes in the underlying algorithm.
Further, it is important to note that, despite our reference to ``Explainable AI,'' ADM is not constrained to AI, and indeed might encompass a broader problem space. 

Despite the emphasis on "understanding" in XAI research, the field features only a few studies that introduce learning frameworks from other disciplines. We argue that interdisciplinary approaches, which can combine the technical explanations of interpretability with more human-centred aspects of HCI or algorithmic education, can produce explanation designs that are easier to adapt to different learning goals and are thus more effective in creating stakeholder understanding. In the following, we briefly motivate the importance of explaining "high-risk" ADM systems, give a short overview of crossovers between XAI and other disciplines to create understanding, and present our approach of incorporating theory from the learning sciences in a qualitative task-based study.

\section{Interdisciplinary XAI approaches to create understanding}

%Compared to AI literacy or AI education, XAI tends to have a lesser focus on classroom settings and instead examines the effects of specific explanation techniques on e.g., lay people and domain experts. However, potential overlaps to AI education and literacy are rarely discussed. 
As mentioned, few studies in XAI actively include frameworks from other disciplines to guide their research, in contrast to works in algorithmic education and literacy. We briefly summarise three such approaches.

Kawakami et al.~\cite{kawakami_towards_2022} describe how Wiggins' \& McTighes'~\cite{wiggins_understanding_2005} "understanding by design" framework can be used to define learning goals together with stakeholders, address all "facets of understanding" and convey the "big ideas" of AI -- an approach that we coincidentally pursued in our study. Kaur et al.~\cite{kaur_sensible_2022} operationalise "sensemaking theory" and state that explanations should take into account "the individual, environmental, social, and organizational context that affects human understanding." Lastly, Alizadeh et al.~\cite{alizadeh_i_2021} leverage the concept of "folk theories" to examine how individual assumptions and expectations are helpful to "discover more valid explanation use cases"~\cite{alizadeh_i_2021}. %These approaches point to the necessity of building explanations that are human-centred, considering individual domain knowledge and sociocultural context, as this affects how people understand and care about explanations.

As of yet, few empirical XAI studies leverage these interdisciplinary findings, although previous work by Miller~\cite{miller_explanation_2019} has demonstrated just how fruitful this can be. In contrast, algorithmic literacy and education studies frequently use established theories from, e.g., the learning sciences~\cite{long_role_2021-1, doroudi_towards_nodate}, with success. %An exchange about teaching methods suitable for different learning settings might thus benefit less human-centred explanation designs. 
To test such an adaptation of "learner-centred" methods to explanations of algorithmic decision-making, we conducted an empirical study, which we briefly describe in the next section. % with 30 participants on the topic of public employment. We applied the "six facets of understanding"~\cite{wiggins_understanding_2005} framework to examine participants' understanding after receiving an explanation.   

\begin{comment}
\begin{itemize}
    \item Short overview of XAI - learning sciences crossovers
    \item 1: Kawakami et al.: Centering XAI towards learning goals, using Wiggins \& McTighe
    \item 2: Kaur: Introducing Sensemaking theory focussing on who the explanation is intended for
    \item 3: Alizadeh et al. and Maitz et al.: Folk theories 
    \item But: The most work is done in education for children, not for adults
    \item And: There seem to be very little \textbf{empirical} studies on how to infuse explanations of ADM with learning sciences insights  
\end{itemize}
\end{comment}

\section{Application of "six facets of understanding" framework in our case study}

\begin{figure}[]
    \centering
    \includegraphics[width=400px]{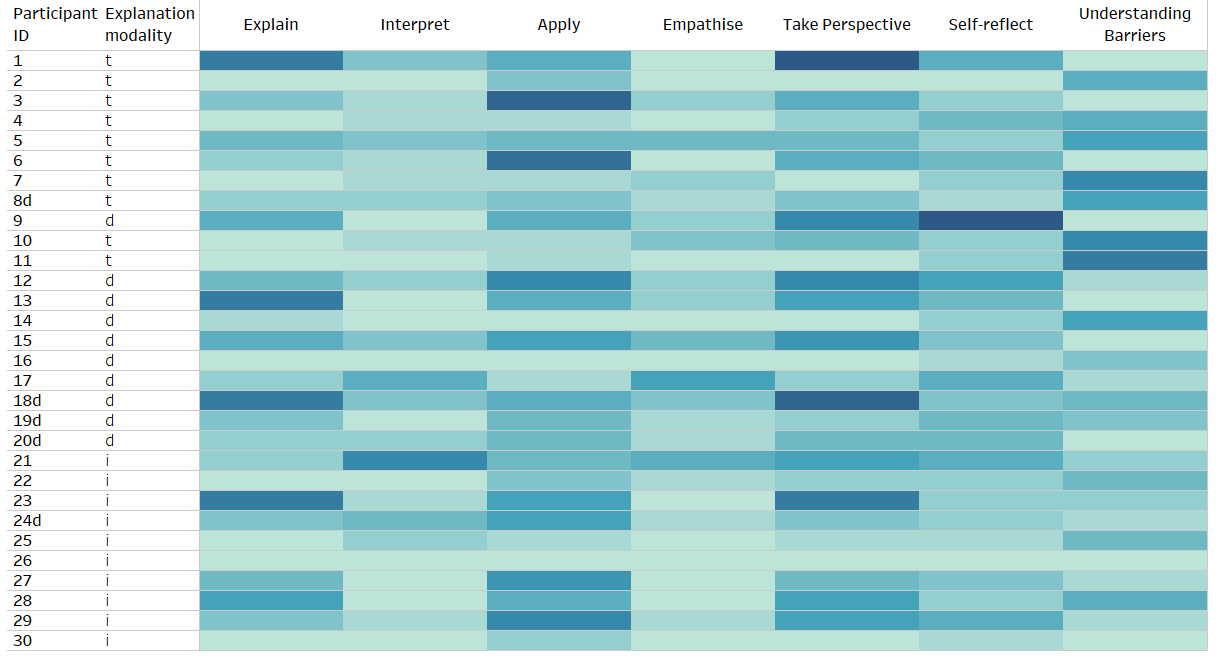}
    \caption[Heatmap of understanding facets]{Heatmap of understanding facets that emerged in participants' responses after receiving one of the three explanation modalities: (t)extual, (d)ialogue, (i)nteractive. Darker blue indicates more frequent use of a facet. The column "barriers" depicts the barriers to understanding a participant encountered. Domain experts are indicated with a \textit{d} attached to their ID.}
    \label{fig:heatmap}
\end{figure}

We conducted a qualitative task-based study with 30 participants on public employment, using the "six facets of understanding"~\cite{wiggins_understanding_2005} framework to analyse participants' understanding after receiving an explanation. The full description and analysis of our study are publicly available~\cite{schmude2023}.~\footnote{Counts of qualitative codes were omitted on purpose~\cite{sandelowski_real_2001}.} Our participant sample included the stakeholder groups "users" (individuals who would use the system as a tool in their work) and "affected individuals" (the "decision targets" of the system). We provided participants with three explanation "modalities" (textual, dialogue, and interactive presentation) of the \amsalgorithm\footnote{The abbreviation AMS stands for the Public Employment Agency.}---an ADM system planned to predict job-seekers' employability in Austria, which was stopped before its implementation. %All studies were conducted in one-on-one interview settings, which allowed for detailed data on participants' understanding and algorithmic fairness perceptions. 
We inductively and deductively analysed responses, using Wiggins' and McTighe's~\cite{wiggins_understanding_2005} framework of the "six facets of understanding," which states that when learners understand, they can: explain, interpret, apply, empathise, take perspective, and self-reflect. An overview of the emergence of facets of understanding is depicted in \autoref{fig:heatmap}. We further examined participants' fairness assessments as an additional indicator for their understanding, following Langer et al.~\cite{langer_what_2021}. 

We found that the "six facets of understanding" framework had multiple advantages in analysing participants' responses: i) it highlighted "emotional" facets such as "empathise" and "interpret," which are not always considered in XAI studies compared to the more "analytical" facets, ii) it allowed us to analyse when participants combined multiple facets in a single answer, showing higher understanding, iii) it includes reflection and "metacognition" as valuable components in understanding, and iv) it includes ways to categorise barriers to understanding, such as forgetting or difficulties to apply learned information. However, we also encountered challenges in applying the framework, as we i) could not easily distinguish between participants' understanding of the domain (public employment) and the algorithm, which points to the interconnection between system and sociotechnical context~\cite{dhanorkar_who_2021}, and ii) encountered strong emotional statements by participants that we could not capture with the "six facets".  

\setlength{\belowcaptionskip}{-15pt}

\begin{comment}
\begin{itemize}
    \item We did such an empirical study: Here is what we found
    \item Framework is applicable to the analysis of understanding (Scatterplot Matrix)
    \item And it helps evaluate the success of certain modalities (dialogue <-> textual, multiple facets = higher understanding, reflection)
    \item But separating understanding domains can be difficult (public employment <-> algorithm)
    \item And systems can be hardly separated from their sociotechnical context, which impacts the meaning of "understanding" significantly 
\end{itemize}
\end{comment}

\section{Open questions}

Our findings show that applying learning sciences frameworks to XAI contexts can provide new perspectives to analyse study data and, consequently, inform explanation design. However, applying these frameworks to a study about "high-risk" ADMs requires a "translation" from their original educational context, as target audiences, available resources, learning objectives, and processes might differ. 

The questions of how this "translation" can succeed and how the evaluation of an ADM system's ethical values are embedded in these frameworks would thus be suitable topics for the workshop. We see a challenge in that XAI target audiences are stakeholders whose day-to-day work does not necessarily include learning activities and who often have little time to spend or are restricted by organisational constraints. Nevertheless, these individuals are in positions where they decide to use "high-risk" ADM systems and thus bear plenty of responsibility. Introducing more learner-centred frameworks to XAI could enable stakeholders to better meet this responsibility by providing them with explanations that increase understanding more effectively.

\begin{comment}
\begin{itemize}
    \item So which learning sciences frameworks are suitable to use in practice?
    \item And how can we adapt them for different contexts and different stakeholders, how can they effectively guide explanation design?
    \item Challenge: K12 education is large-scale, students are taught over the course of a dozen years. Explanations for adult people need to be brief, concise, flexible, and relevant for their field of practice, without oversimplifying.  How's that possible?
\end{itemize}
\end{comment}

\begin{acks}
This work has been funded by the Vienna Science and Technology Fund (WWTF) [10.47379/ICT20058] as well as [10.47379/ICT20065].
\end{acks}

\bibliographystyle{ACM-Reference-Format}
\bibliography{references}

%\appendix

%\include{source/supplemental material.tex}

\end{document}